# Supporting the Use of User Generated Content in Journalistic Practice

**Peter Tolmie**
University of Warwick
Coventry, UK
peter.tolmie@gmail.com

**Rob Procter**
Warwick University
Coventry, UK
rob.procter@warwick.ac.uk

**David William Randall**
University of Siegen
Siegen, Germany
daverandall2008@gmail.com

**Mark Rouncefield**
Lancaster University
Lancaster, UK
m.rouncefield@lancs.ac.uk

**Christian Burger**
Swissinfo.ch
Bern, Switzerland
Christian.Burger@swissinfo.ch

**Geraldine Wong Sak Hoi**
Swissinfo.ch
Bern, Switzerland
geraldine.wongsakhoi@swissinfo.ch

**Arkaitz Zubiaga**
Warwick University
Coventry, UK
arkaitz.zubiaga@gmail.com

**Maria Liakata**
Warwick University
Coventry, UK
m.liakata@warwick.ac.uk

## ABSTRACT
Social media and user-generated content (UGC) are increasingly important features of journalistic work in a number of different ways. However, their use presents major challenges, not least because information posted on social media is not always reliable and therefore its veracity needs to be checked before it can be considered as fit for use in the reporting of news. We report on the results of a series of in-depth ethnographic studies of journalist work practices undertaken as part of the requirements gathering for a prototype of a social media verification 'dashboard' and its subsequent evaluation. We conclude with some reflections upon the broader implications of our findings for the design of tools to support journalistic work.

## ACM Classification Keywords
H.5.3. Information interfaces and presentation (e.g., HCI): Group and Organization Interfaces; Web-based interaction.

## INTRODUCTION
Social media and user-generated content (UGC) are increasingly important features of journalistic work in a number of different ways [11, 13, 24, 34, 36, 51]. They are used as resources for leads and the identification of stories [18]; as sources of content and to facilitate the verification of content that has itself appeared on social media [60].

And they are also used in their own right as a vehicle for publishing news [27]. It is therefore unsurprising that use of UGC is now strongly embedded in routine everyday practice for many journalists. A broad swathe of literature also recognizes how it has become a pervasive feature of the work [19, 39, 51]. However, its use also presents major challenges, not least because information posted on social media is not always reliable. Its veracity therefore needs to be checked before it is used to support a news story [7]. Indeed, numerous sources attest to how this is now a major concern within the industry [13, 24, 50]. In this paper we report on a European research project where journalistic practice was initially studied ethnographically. Results were then used to develop a journalist dashboard to support aspects of journalistic work. We present findings from evaluations of that dashboard and discuss the implications for future design of tools to support various current and emergent journalistic practices around the use of UGC.

The central interest of the Pheme project [41] is the use of machine learning to provide assistance in the determination of the veracity of rumors in social media [7,60]. The aim is to provide journalists with a better view of what might constitute a trustworthy piece of information and to understand how rumours emerge and unfold.

A range of tools shares certain commonalities with the Pheme journalist dashboard. FactWatcher [25] offers a dashboard presentation by which journalists can identify three types of facts with associated data that could serve as leads for news stories. RumorLens [46] provides a dashboard type tool for visual analysis of data that has been mined from Twitter. This aims to help journalists identify rumors as they arise and to assess their spread and the extent to which they have been subject to correction. TwitterTrails [20] is a web-based tool. Like RumorLens it can be used to examine how a rumor first arose on Twitter, how it spread, and the ways in which it might have been refuted. On TwitterTrails, journalists can take keywords from a specific tweet and use them to drill into its origins and other related tweets. On a different tack, CommentIQ provides a way of identifying usable UGC in terms of online comments from readers of already published news articles [37]. Yet another approach can be found in visual document mining tools such as Overview [10].

Vox Civitas [18] was designed to assist journalists in the location of newsworthy stories amongst the vast quantities of social media content available. A dashboard-type design enabled users to play video clips of broadcast media news events. Alongside of it they could view a stream of twitter messages associated with those clips. Graph-form analytics then displayed features such as the tweet volume over time, associated levels of sentiment/controversy, and the keywords present at any selected point on a timeline. Some of the same researchers [19] were involved in the later design of SPSR (Seriously Rapid Source Review). This offered a number of the features built into the Pheme journalist dashboard, including search, sort and filter functionality, the tweetstream relating to specific events, the author's location and details, keywords, and associated features such as images and video. This work is amongst the most closely aligned to our own but does not scope its emphasis around verification and veracity to the same degree. The Reveal project [47] has been similarly concerned with providing tools to support journalistic work in the context of the propagation of false information and rumors. Media REVEALr [3] does provide certain kinds of verification support. This includes the detection of duplicate images against an indexed collection, grouping of posts by their similarity, and grouping and visualization of specific entities associated with an event.

The Pheme journalist dashboard is unique in that it provides journalists with computational assistance in the form of machine learning algorithms trained on conversational tweet threads [54, 60. These are capable of assigning automatically a veracity score that reflects the dashboard's confidence in a given rumor being true at any given time as the story unfolds. Initial training data here has involved crowdsourced annotation of related tweets to identify how source tweets align with or refute particular potentially rumorous news items. Annotation of subsequent tweets focuses on the extent to which they agree or disagree with the source tweet. Across all tweets further annotation indicates the degree of certainty present within each post. In all cases annotation was conducted over the whole span of particular twitter conversations in order to identify temporal patterns in how rumors unfold. To the best of our knowledge in earlier dashboards assessment of rumor veracity remains essentially a manual process.

Initially, we conducted detailed ethnographic studies of newsroom environments where UGC is a feature of the journalistic workflow. The workflow and associated journalistic practices were explicated in order to understand the use of UGC in the broader context of everyday journalistic work. Analysis of these materials provided us with a body of key requirements to be taken into account when designing tools for journalists to use when handling UGC in newsrooms. This was followed by a series of iterations and evaluations of dashboard prototypes. These were designed to support the incorporation of UGC into the journalistic workflow, and the timely identification of rumors and their likely veracity.

In this paper we focus upon the ways in which the observations and evaluations have surfaced some important issues regarding the use of UGC in journalism. UGC is by nature rapidly evolving, often large in volume, and involving multiple sources of unknown reputation. Our findings therefore focus on how UGC is verified to meet traditionally exacting terms of journalistic probity.

In order to articulate these matters more fully we draw on well-documented, key characteristics of newsroom activity [9, 38]: the temporal fluctuations in how much pressure journalists are operating under; relatedly, the presence of cyclical patterns in the work; the shaping influence of deadlines; the varying (and sometimes contradictory) organizational imperatives that have to be attended to (e.g., the need to build readership and satisfy advertisers by delivering large numbers of page impressions); the various national and legal frameworks within which journalists have to operate (e.g., libel laws, privacy); the varying exercise of editorial control; the local environment in which news stories are constructed; the interests of the audience being addressed; the different formats being worked to, which can range from short news updates to features and articles; and, of course, technologies and software used for both sourcing and publishing stories. Notable in all of this is the way in which just what UGC might relate to, and the way this is arrived at in the course of everyday working practices, cannot be divorced from the organizational and social context in which the news is getting produced.

We conclude with some reflections upon the broader implications of our findings for the design of tools to support journalistic work involving UGC.

## JOURNALISM AND UGC

Journalists have long used information and communication technologies (ICTs) in their work. Computer-assisted reporting (CAR) was first used in the 1950s to support journalists covering election results [14]. More recently, the rise of 'data journalism' has signalled increasing reliance on publically available datasets to provide the evidence and analysis to support news stories [17].

Most recent research into journalistic practices has focused mainly on the way in which UGC, the rise of 'produsers' [12], social news and citizen journalism, is changing the way news is gathered, analysed, reported, and disseminated by online and traditional news media [11, 34, 36, 42, 51]. An important aspect of this has been convergence between social and broadcast media in the newsroom [23]. The integration of UGC into newsroom routines has become key. This is especially so where there are limits to media access. This may be imposed by government authorities, e.g., the Iran protests and Arab Spring uprisings; or natural, e.g., disasters such as Hurricane Sandy. It is also key where events are unfolding at a fast pace and/or professional journalists are not yet on the ground [22, 24].

Journalists now inhabit a world where information is abundant but where traditional methods for determining its relevance and quality for reporting have not kept pace. Such trends cut across journalistic ideologies in respect of knowledge and expertise, and concern for balance and verifiability. These are what Zelizer [58] refers to as 'god-terms'. The principles of verification are, in this view, problematised by the new media, and by UGC in particular, especially in the context of fast-breaking stories. This, Hermida [28] has argued, is leading to a move towards 'collaborative verification'. "*The acceleration of the news cycle and proliferation of news and information has raised concerns about the erosion of the discipline of verification, and by implication, the professional legitimacy of journalism… Major news organizations... have published accounts of breaking news events in ''live updates'' pages that combine unverified social media content and authenticated professional reports.*"

Much of the above concerns the way in which journalists orient to and use new online services that complement, or sometimes challenge more well established news organisations. Facebook has also launched 'Signal', a tool for journalists to access and use trending events on Facebook. And it has become commonplace for journalists to make use of Twitter as one of a battery of social media resources they monitor for leads on a regular basis [59]. Although such developments mean that, more or less by definition, newspapers and television content producers will have to negotiate a different relationship with their audience, it is also the case that 'citizen journalism' can, in principle, constitute a wholly alternative source of news [32]. This can entail acting, for instance, as a portal for other news outlets, blogging and peer-to-peer opinion exchange. There are many such examples, often coming with a variety of political stances [1, 2, 48, 55]. Traditional news organisations face a challenge to maintain their reputation for reliability of reporting while demonstrating they can keep up with a news cycle that seems increasingly driven by UGC. To this end many have built dedicated UGC handling units into their organisation [13, 23, 30, 32].

The Pew Research Center [40] reports that between 2013 and 2015 the proportion of new media users relying on Twitter and Facebook as a primary source for news increased from 52% to 63% in the case of Twitter and from 47% to 63% in the case of Facebook. Moreover, and perhaps of more consequence, 59% of the former use it as a means to keep up to date with unfolding events. Twitter, in other words, seems to be a significant source of 'breaking news'. The launch of 'Moments' on Twitter is an explicit response to this demand [35]. Such moves have proven controversial. While it can be argued that 'citizen journalism' is a democratising tendency, concern has been expressed over its reliability [6, 16]. There have been examples of inaccurate information indiscriminately shared resulting in risks to public safety [44], harassment [53], violations of social media platform policy [52] and law [5].

Previous studies of journalistic practices have highlighted the importance of collaboration for production of stories to tight deadlines while meeting professional standards for accuracy and quality [26, 31]. This remains just as important in the modern newsroom with the advent of UGC and the 24/7 news cycle because the workflow typically involves contributions from different newsroom participants. It is a series of activities that begins with identification of candidate stories and ends with publication of those that survive the selection process. It may be tempting to seek to formalise these steps, but studies of the use of workflow systems in practice reveal time and again that such simplifications often hinder people's capacity to 'get the job done' if implemented in that fashion (e.g. [8]).

### INITIAL OBSERVATIONS OF JOURNALISTIC PRACTICE

As a first step towards understanding how use of UGC is embedded within journalistic practice, we undertook an in-depth ethnographic study in a national news organisation in Switzerland that produces exclusively online content. This news organisation is championing more thoroughgoing exploitation of UGC and has closely modelled much of its approach upon practices developed by the BBC [4].

**The Study**

The ethnographic study was conducted over two separate blocks in April and November 2014, each lasting for a whole week. The aim of the study was to get a proper situated understanding of the real nature of the work. The focus was therefore on close observation of the actual ongoing work of journalists on the newsdesk. 15 journalists were observed in this way over the course of the study, each for between 2 hours and half a day, including editorial staff. The work was systematically recorded across the whole study using a mixture of video and audio recordings and handwritten notes. All of the data was later analysed closely to pull out the constitutive practices of the work and to explicate the methodical ways in which these practices were accomplished in ongoing action and interaction. This ethnomethodological approach has become one of the mainstays of ethnographic work in HCI [15].

A range of practices was observed across the workflow for story production, editorial review and publication. This included use of UGC to a variety of ends and included an array of activities involved in verifying story details. Verification and the use of social media were also discussed in further detail in several dedicated interviews.

**Findings**

A primary outcome of the ethnographic work was a full mapping of the journalistic workflow and the many contingent elements involved in its accomplishment. In outline this amounts to:

*Looking for stories:* Prospecting for raw materials that might be turned into stories is a critical part of journalistic work. First thing in the morning, whenever it is quiet, and between other ongoing tasks, journalists will recurrently monitor a range of resources that might provide them with leads, such as news wires, social media (especially Twitter), other news sites, and their email.

*Selecting stories:* This is particularly challenging. First, a potential story has to be recognised. Here, a range of considerations may be brought to bear, e.g.: Does it fit the organisational remit (most news organisations have criteria that govern what kinds of content should be covered)?; Is it news?; Is it interesting enough to merit publication?; Is it a busy or a quiet news day (on quiet news days stories may be considered that would be rejected when it is busier)?; and so on. Once a potential story has been spotted, any decision on its viability generally relies on input of other newsroom participants. These may include colleagues and editors, and may get handled through informal conversations or during meetings. Other people may often propose angles worth pursuing and editors, in particular, may ask for further evidence and validation.

Journalists will typically have a number of potential stories they are looking at, so there is also a moment where they select one particular story to start developing. Which one gets worked up will depend on matters such as: timeliness; presence of adequate evidence; time of day (journalists adhere to deadlines and different daily content cycles); etc. Throughout there is a continual process of checking and verifying information.. This will make use of: web-based resources and searches; official documents where they can be turned up; interviews with 'knowledgeable' parties (witnesses, experts, etc.); and so on. The combination of all these different factors means that story selection is actually a complex calculus for journalists that has to be worked anew in every single case. It is not something that can be provided by the simple 'cranking of a handle'.

*Actually writing the story:* It was recurrently observed that journalists begin by opening a blank Word document, into which they paste all the materials that might be relevant to the story from various places, including wires, tweets, passages from web documents, emails, and so on. They then assemble the actual text in reference to these, above or beneath the copied material. The process of writing up here is one of the primary prompts to further checking and verification as specific details are encountered. Additional 'completion' activities include: checking word lengths to bring it within specified constraints; providing the story with a lead; a title and surtitles; and the location or creation of additional elements such as pictures and infographics.

*Getting the story 'subbed':* It is standard practice that another journalist will 'sub' a story by reading it through and undertaking corrections and edits to ensure it is 'up to scratch'. This stage will often implicate further checking of specific facts and figures as a means of quality control.

*Publishing the story:* Once a story has been subbed it is usually published. Publication frequently involves visual checks and assignation of metadata to stories, e.g. category labels, IPTC codes (a set of agreed controlled vocabularies for news stories [29]) and links to related materials.

It should be noted that even beyond initial publication there can be further phases of review, verification and revision as stories are checked over by editorial staff. This underscores another key finding that came out of the ethnographic work: checking and verification activities are visible *throughout* the workflow. Verification is not a 'one stop shop', with it happening at just one dedicated moment. Instead it is a concern that becomes relevant again and again throughout the process. So tools to support verification work need to support its *ongoing* character, including making visible what verification work has already been undertaken.

| *Information need* | *Interaction need* |
|---|---|
| **Uncovering stories** | |
| Trending information/topics | Save / mark for later |
| Last update (or real time) | Read saved / marked information (how long back?) |
| **Selecting stories** | |
| Veracity indicator | Mark for verification |
| Verified information (sources, entities) | |
| **Handling supposition, speculation and rumor** | |
| See if information is supposition, speculation or rumor | Review steps that have been taken in verification process |
| Level of veracity | |
| **Verification** | |
| Marked items or information for verification | |
| Verification hints:<br>- location<br>- history (if available) | Associate verification process to specific item / information |
| Verification sources (e.g. contacts in house / external) | |
| **Revision and re-verification** | |
| | Integrate personal UGC into tool |
| Provide views for quick overviews (summaries) and detail-views for in-depth understanding | Allow cross-personal working: sharing of all information on specific tasks |
| **Practices** | |
| Keeping in view relevant resources attached to both personal and organizational identities (e.g. Twitter accounts) | Selecting, collating, switching between relevant resources attached to both personal and organizational identities |

**Table 1: Core Requirements.**

**Initial Requirements**

Requirements arising from the ethnographic work were broken down into a core set of information and interaction needs relating to the identified journalistic work processes [45] (see Table 1).

**THE JOURNALIST DASHBOARD**

The findings from the ethnographic work were then used to inform the iterative development of a journalist dashboard, which began as a series of wireframe mockups but which had become a working prototype by the beginning of 2016. All versions of the dashboard focused upon the capture and presentation of materials from Twitter as this is the primary source of UGC for news desk activity.

**The Dashboard**

The initial version provided for a range of functions. These included presentation of tweets grouped together in 'stories'; a conversation (i.e. a series of tweets where successor tweets are replies to predecessors) 'history' showing how tweets appeared over time; details on tweet authors; annotations regarding the veracity of the various tweets; a map to allow journalists to localize tweets and their origin; associated reports that had been linked to in the original tweets, such as further news reports; and a placeholder page for where journalists might ultimately also have access to various images and videos associated with the various tweets presented. However, there was as yet no provision for direct interaction with live Twitter data and limited scope for searching or filtering. Evaluation led to a further iteration where users now had access to 'live' data.

Evaluation of this second version, in turn, surfaced a further set of critical issues, which are now in the course of being addressed. Latest versions of the dashboard provide a full range of functionality except for certain additional display capacities (such as extraction of detail provided by hover-over in the original Twitter interface). Evaluation of this will happen later, but our concern here is not to present a technical discussion of the dashboard but rather to explore the broader implications for *all* such endeavours surfaced by our study and the evaluations we have undertaken.

**EVALUATION**

Evaluation of the initial, non-live version of the dashboard took place in January 2016. It involved 2 journalists from the same organisation we had studied ethnographically. They were chosen because they used UGC in distinct ways. It used canned data from an annotated dataset that mostly related to the Germanwings plane crash in March 2015. It pursued an instructed process mimicking the basic journalistic workflow and working between newswires, the dashboard, associated news reports and images, and a Word document in which to copy uncovered information. The journalists were asked to use the dashboard as though they were looking for a story to prepare.

Evaluation of the 'live' version took place in April 2016 using the same 2 journalists. Here the journalists attempted to use live data that was related to the pre-specified search term, '(Donald) Trump'. Aside from this the intention was to repeat the same basic process as the first evaluation, with journalists preparing a 'mock' story on the basis of materials uncovered.

**Findings**

Initial evaluation uncovered the following matters:

*Relation to Journalistic Workflow & Resources:* Dashboards like this need to be able to support two distinct use cases. 1) Working on the *newsdesk*, where journalists are less interested in conversation history, origins, and tweets that the system has already identified to be rumourous and untrue. 2) Writing *features*, where the conversation history, origins, and potential sources of rumour *are* of interest. The difference here arises because an overriding feature of work on a newsdesk is pressure of time. When time is tight journalists need to know the exact state of affairs right now. This has implications for the liveness of data and the frequency of updates. With feature writing journalists have more time to work on their stories, sometimes days, weeks, or occasionally even months. This allows journalists to drill into the available resources in greater depth. Thus, the dashboard was considered to be a useful adjunct to the wires and a potential quick route to associated reports. It was also more broadly seen to offer to-hand support for story justification when proposing stories to editors and to provide a means of uncovering details about a topic that might not otherwise be uncovered.

*Features:* A key point was the presence of a map, seen as being potentially of high value if it could rapidly display where tweets are coming from and to select a subset of tweets and drill down. Journalists would also like to see author information attached to tweet labels on the map so that they can quickly assess the likely significance of the author. Author information needs to be available because it is one of the key criteria used for assessing the interest of a particular tweet. There is also a strong interest amongst journalists in being able to see tweets according to event proximity. It was also noted that journalists needed to be able to see information about people who are mentioned *within* tweets, perhaps by using a mouse-over display.

Another feature deemed to be of definite value was the access the dashboard could give to associated reports, which journalists would also like to see localized. Related pictures and videos captured from the tweets were also considered useful, both as a possible source of additional detail, and as a possible source of actual images to use.

The potential value of having a conversation history, however, was less clear-cut. It was felt to be of most use when there is uncertainty about a topic. It was also seen to be potentially useful for identifying the source of a rumour, and potential witnesses and experts. Time constraints, however, made its use more evident for features.

*Noted Requirements:* The evaluation also surfaced elements that were not currently available. Being able to sort and filter was particularly emphasized. This is a key requirement, with two parts: 1) the shaping of the landing page with pre-specified filters, much as they can already do with the wires; 2) being able to apply additional filter and sorting options. Sort functionality should cover the following elements: Time Order; Veracity; Tweeter Location in Relation to an Event; Number of Followers; Language; Number of Contributions to the Tweet Thread; speed of the development of a rumour. The preferred defaults for sorting are time order or veracity. However, there was some variation between the journalists on this, emphasizing the point we shall be making later that requirements for dashboards can be hugely contingent and context dependent. Journalists would also like to be able to sort within already filtered lists. Having search functionality, however, is the most important requirement of all. Journalists need to be able to search on compound as well as single terms and searching needs to return results quickly for journalists working on the newsdesk.

For the purposes of veracity assessment journalists would like to have general confidence measures of how accurate tweet content might be, together with a display of changes in confidence over time. A graphical display is preferred. They also need to be able to organize tweets in relation to trusted sources, i.e. other trustworthy news organizations. Additionally, foregrounding of factual content (names, numbers, specific facts, etc) would be useful, together with a capacity to sort accordingly. They would also like to be able to inspect the *grounds* of any veracity assessments the system has made upon request, rather than simply trusting that the technology has got it right.

Additional features of importance are whether tweets have been retweeted, favourited, etc. and trends in tweet/thread popularity over time. Once again, a graphical display would be preferred. One other noted requirement was being able to preserve a use trail. For this they would like to be able to mark-up specific tweets in the dashboard (for instance, by starring them). They would also like to be able to build collections of specific sets of tweets that they can return to later for justification, verification, etc. and be able to share these with other people, such as editors.

A second evaluation revealed additional matters:

*Lack of 'liveness':* The system had been set up so that 'live' data should have been visible on the dashboard. However, during the evaluation the refresh was patchy. The journalists started out with data 11 hours old, and even after an update happened, it quickly became out of date. This made clear the need journalists have for up-to-the-minute information, with older data causing notable problems.

*Search:* At this point there was still no search option. The fact that there was no search or filtering functionality meant that the stream was undifferentiated, with the journalists having to scroll up and down extensively.

*Coherence:* Keywords associated with a particular cluster in this version were considered unhelpful. The tweets were grouped together according to a clustering algorithm but the clustering criteria were opaque to the journalists and the clusters therefore confusing. Furthermore, the pre-defined term. 'Trump', was highly restrictive. They were not able to refine it by adding words to it and it did not relate to a topic they would typically write to.

*Temporal cohesion:* On top of this, it was not clear that the grouping of tweets to conversational threads was happening properly. It seemed to be the case that tweets falling outside of a certain time window were not necessarily treated as part of the same conversation. It also seemed possible that conversational threads would not be preserved over time. This was a particularly worrying issue because being able to see conversational threads is core to being able to identify unfolding rumours and their spread (see [60]).

Many of the above issues have been addressed and a new version of the dashboard will be evaluated in September 2016. We wish to draw upon the insights we have collectively gleaned from the ethnographic work and the evaluations undertaken so far, identifying important lessons that speak generally to the design of these kinds of tools.

**ANALYSIS AND DISCUSSION**

A key contribution here is the depth with which accomplishment of everyday work in newsrooms was investigated. This produced insights that are hard to gain in any other way. There are two sets of insights that have come out of this that we wish to emphasize in particular here. One of these is *the highly contingent character of newsdesk work*, the other is *the extent to which verification is an ongoing and contingent process* as well. In the following discussion we will tackle these issues conjointly.

*Time pressure, temporal cycles, and deadlines:* One of the strongest characteristics of newsdesk work is the routine oscillation between periods of intense time pressure and periods of relative calm (see also [19] regarding this). As this is a product of external events that are completely outside of the control of the journalists themselves there can be days on end when the pressure is continuous and other days when it is hard to find anything worth writing about. This has implications for what may usefully be drawn upon in the way of UGC, regarding the performance of a system and its capacity to deliver content that is genuinely 'live' and up-to-the-minute. This was forcibly articulated by the journalists involved in our evaluations:

**Ethnographer**: Would 10 minutes be deal breaker for you?
**Journalist:** Well, it depends what it'll do. If it's going to evaluate and process and filter things and do some of the work for you that might help. Then I would give it a bit longer ... But you could look on Twitter for yourself in 10 minutes… find some people… click and see where a person is if you want .. I can't say until I see what it comes back with … but I definitely know that 10 minutes is too long

A consequence of all this is that when a newsdesk is busy results have to be found fast and threads are dipped into opportunistically to provide possible interesting leads and angles or specific details that have not been reported elsewhere. Conversational threads are not examined systematically from their inception because there simply isn't time. On a quiet news day, by contrast, it may be possible to explore specific conversations more thoroughly and even to pull together whole bodies of UGC to engage in various kinds of meta-analysis of how events have unfolded. This, too, was articulated in one evaluation:

It depends what time you're doing it at as well … If it happened at 10am and you were writing at 11 then you wouldn't go so far back to look at things that happened a while ago. It would probably be resolved if it was true or not by that point … But I might be writing a feature, not breaking news. For instance how it happened. Maybe a scandal or something. What someone said first and how they got corrected. I would be interested in going back further in the conversation history then.

An obvious design implication of all this is that *no single presentation of UGC is going to suffice*. Instead the presentation needs to be adaptive to circumstance, with different modes being easily toggled by the journalists themselves. A typical response to this requirement in dashboards is to provide a landing page that gives a broad overview of currently available content, with an option to drill into different features according to need. However, this still implies choices regarding just what is initially presented and drilling down may be tractable to different degrees. The capacity of dashboards to deliver live and comprehensive content is also a challenge because many of the major social media platforms place constraints upon the access third parties may get to their data.

Verification is also a clear issue when time pressures are variable. Something that can provide indications of veracity of UGC in a trustworthy fashion at a glance is clearly desirable when newsdesks are busy. Manual verification and non-automated fact-checking can potentially involve journalists in extensive research and may simply not be feasible on busy days. This gives journalists a strong steer towards using UGC that has itself been produced by already 'trusted' authors, such as other news organizations, so that some degree of verification can already be assumed.

As well as there being entirely unpredictable time pressures in newsdesk work, there are also more stable temporal cycles. Some of these are bound up with things like shift work in order to maximize capacity to cover breaking news as it actually breaks. There are also particular kinds of stories that are often run at particular times of day, for instance sports news in the evening and financial news in the morning. News cycles may also get set up around planned national and international events. Anticipated major events can have especially strong implications for how news organizations handle UGC:

[Journalist talking about preparations for a national vote] I'll use Storify to assemble it. That way I can share stuff on our main page. That's what people usually look at. For the Twitter feed I can keep reusing the same link so it's quite efficient. I tried Storify before for a government election. I was the first one to do it that way. I got a lot of people following me because of that … But you need the right monitoring tool for finding out what's being said. Even on Twitter you need the right list. So you have to second guess who might be talking about a particular topic … And T *[a colleague]* is doing some prepared tweets, having things ready that he can post when it's time. You can't do it all on the hoof if you're going to do it properly

Knowledge of up-and-coming events may also shape what kinds of UGC is looked for on a daily basis:

We get a daily programme - it's usually about a quarter to 9 they start coming in - that'll tell you the things that are happening, and what the wires will be filing on throughout the day.

Here we can see that temporal cycles of various kinds can have significant implications for both how journalists go about exploring UGC content and how they may approach the direct incorporation of UGC into production of news.

In many instances the design pullout might be that there are certain ways in which journalists are already anticipating the use of UGC as a resource and that the thing to support might be the planning of that use. This has to be offset against the fact that newsdesk work is also replete with unplanned use of UGC, so any design is going to have to support both of these extremes.

Verification of UGC in the kinds of situations outlined above also presents some unique challenges in that news production is typically looking to incorporate UGC directly in order to illustrate a range of possible views. Thus the concern becomes one of framing such that readers understand they may be engaging with unverified raw content and this is not something for which the news organization would wish to be held accountable. Underlying this is a further issue of selection with regard to unverified UGC. The very act of incorporating UGC such as individual tweets may be sufficient to promote that tweet. There are already a number of cases where news organizations have been the unwitting disseminators of false rumors in this way [33, 43].

*Organizational imperatives:* A key problem confronting designers of tools for journalists is contending with the fact that a wide variety of organizational imperatives, constraints, and styles may be encountered across the sector. Each of these may have significant impact upon how the right kind of support might be provided (see also [37] on this point). One important way in which this may be articulated is through what a specific organization may view as viable content for its publications. Here a journalist outlines how he has set up filters on his newswires to specifically accommodate these constraints:

We've all got our own filters, which, there's er, for me, I'm guessing about thirty forty words that we all have, like Schweiz, Suisse, Bundesrat, Senate, House of Reps, All the seven ministers' names, … all to do with Switzerland …

This places strong requirements upon the design of resources that furnish UGC to journalists because there is little point in providing something that cannot be shaped to need in the same way as other resources like the newswires. This means at a minimum that tools will have to provide good search functionality and a capacity to apply filters and sort content and this proved to be one of the core concerns of journalists when confronted with a new tool.

Organizational constraints may also operate at the level of technology provision and this can have an important impact upon how UGC may be used by journalists. One obvious issue here is the kind of technology journalists have available to them in the first place. A notable consequence of this is that journalists may circumvent the limitations of workplace technology by supplementing it with their own devices as well, many of which may be expressly used to furnish them with further leads: *[Journalist looking at Twitter and scrolling down. Her iPad makes a noise]* That's a BBC news notification. I've got one for 20 Minutes (a local publication) as well. It's my own tablet but we do have some office ones as well. I also have RSS feeds going to my iPhone.

If dashboards are put in place as organizationally accredited resources, a question arises as to how to handle the way journalists currently access UGC across multiple devices and may want still dashboard functionality across that ecology of devices, rather than having to relocate already uncovered UGC in a dashboard on their work machine.

A further constraint that can arise is the use of personal and organization-based UGC accounts. Organizational UGC accounts are typically set up to provide feeds from sources that are seen to have cross-organizational relevance and that meet with organizational approval. Most journalists, however, have personal UGC accounts that they have specifically crafted to their own interests with a list of people they follow accumulated over a number of years because of their utility for providing good leads. Thus some journalists we observed kept both their personal and the organizational accounts open simultaneously by running them in separate browsers. However, we also saw many journalists working on machines where they did not have this option available. This tension between personal and organizationally accredited use of UGC also has an impact upon the *posting* of UGC. News organizations use UGC to an increasing degree for both disseminating and promoting their content and the preparation of such content is an increasingly important aspect of journalistic work. However, many journalists also post UGC regularly in a personal capacity. Despite stringent efforts on most of their parts to make it clear when they are posting from a uniquely personal point of view, journalists work against the constant risk of being taken to be posting in an organizational capacity. This can constrain their individual posting.

All of this has additional implications regarding how content is verified. In many instances UGC received on personal accounts is from personally known sources, which can impact on how the veracity of what is being said is understood and the kinds of formal verification steps that are seen to be necessary. UGC from independent sources may require more extensive verification, which can be an issue in fast turnaround situations. Similarly, journalists may not be so motivated to fully verify content they post in a *personal* capacity, leading to some unfortunate outcomes. In so far as they are able to provide a rapid overview of the social media landscape around a specific topic, dashboards may have a role to play in this respect. They may also provide journalists with ways of assessing veracity and checking facts within posts that they might otherwise consider to be too time-consuming. However, there is also a question regarding how far journalists may be able to shape dashboards to meet their own personal preferences, and the extent to which that may or may not be desirable.

*National and legal frameworks:* In many countries, news organizations are required to respect national regulations and industry standards and guidelines that have an impact on what can be reported (e.g., to limit abusive content) and how it may be reported (e.g., to avoid breaches of copyright)". Fundamentally it is understood that journalists have a duty to verify all information they publish. In some countries it is even expressly against the law to knowingly publish false news, but there are variations in the extent to which this may be rigorously adhered. Here a principal editor expresses some of the shaping consequences of this:

In Switzerland people have the right to read [their quotes] before publication, though they can only control what we do in terms of mistakes. For Twitter we use the same verification standards, but there's a problem when things first appear. Once you have the history things are more visible. Often we use experienced colleagues, ask if they are sure, and go ahead if they say yes … But it depends upon the region. With our Arab colleagues it's easier. If you have the history and comments and posts and shares they know very fast. There's a big issue with the Indian press. The problem there is they don't bother to show their sources. It's one of the worst cases … And even experienced journalists have difficulty with manipulating governments, like China and Russia. Then we have to invest a lot of effort and take a different approach … Mostly we trust no-one if we have no arbiters… Quality standards are defined for the private sector. We had to define our own standards as a public organisation … We are quite often criticized for not being balanced, regarding finance, votes, etc. We are aware of it and discuss things with our stakeholders. And we have an ombudsman. If things go to her then we have to account for our decisions.

The specific national context of news production can have a profound impact upon both the perceived need to verify content such as UGC and also the extent to which a national readership may assume that content derived from UGC has been verified. The fact that what may actually matter and be of particular priority in news reporting is not independent of national boundaries means that the very features that are promoted and given preference in dashboards may need to vary according to just where they are being used.

*Editorial control:* A key, and sometimes idiosyncratic, aspect of news production is the role played by editors who usually exercise control over the viability of a story. Some editors are more directive than others but all operate as organizational incumbents to whom journalists are specifically accountable. Knowledge of editor preferences therefore plays an important part in the expectations of journalists regarding how different aspects of stories may be received. Editors may also specifically instruct journalists in a variety of ways. This can have a particular impact upon what kinds of verification may be required.

In the following example we see an editor instruct a journalist regarding exactly what kinds of evidence he requires to accept the story she is working on:

*Journalist and Editor talking about migrant workers piece she is writing. The editor wants numbers included in the piece. They work together through various tables showing numerical information. Then the editor asks her to provide an info box or an infographic. The journalist says she's not skilled in that kind of thing but the editor says that she needs to have something to grab attention.*

**Editor**: Focus on getting comments supported by 2 or 3 people working in that sector. And stress the importance of foreign skilled labour to Switzerland. Find some figures saying 20%, 30% or whatever.

With regard to the use of UGC this may impact on what kinds of evidence criteria it may have to fulfill. Here, for instance, a journalist explains what she needs to bear in mind regarding 'trusted' sources:

Just from what we know and what we are familiar with I would check what Reuters have. They are a trusted source for us. Here they are also confirming 144 passengers and 6 crew on board. So then with that information I would feel comfortable with reporting that figure, since I have from both Reuters and the BBC. But I would continue to keep an eye on it. I would write the initial story but with something like this I'd watch it all day. Eventually the Swiss news agency would put out its own version of events and I would cross-check against. Our policy here is that a fact has to be confirmed by the Swiss news agency or by 2 independent trusted sources.

What this amounts to for the provision of UGC via dashboards is that, whilst dashboards may currently foreground certain elements and de-prioritize others, just what kinds of elements may actually need to be emphasized will be subject not just to the preferences of individual journalists, but also the preferences of specific editors and organizations. This means that dashboards may need to be open to specific user-configuration. Whilst designers might say that if the required features are there journalists will learn how to navigate to them, it needs to be remembered that in newsdesk work time is of the essence and, if dashboards cannot meet the user need quickly, then they will just as quickly be set aside. As one journalist put it when testing an early version of our own dashboard: "I'd already be on Twitter by now."

*The local environment:* News organizations are variable in size as well as intent, and journalists work in a variety of environments including large open plan offices, individual offices, or even at home. Resources available to journalists in terms of equipment, archives, and actual people can vary enormously. All of these factors may impact on how UGC can be appropriately used and verified within the construction of news stories. Furthermore, many journalists may identify potential stories and start to work upon them across a number of different environments, for instance first noticing a story whilst online at home, searching UGC for further details whilst travelling to work, and then working further upon the story once physically at work.

Erm, I mean I get all the- the headlines on my mobile phone. So I already get them on the way in. So I know what - what's to expect and at seven o'clock I listen to the radio .

In the following instance we see how a journalist drew directly upon her peers in the local environment in order to assess the viability of a potential story she had just picked up on when using Twitter:

*C is looking at a wire about guides refusing to take people up Everest. She thinks there must be a Swiss group there and says about someone who has done stories for them who lives in Nepal.*
C: There was a climber who climbs mountains as fast as possible, and the guides were attacking their party and she did an interview with him. I wonder what she's doing?
S: I think she'll be there.
*C looks at Google, then goes to Twitter to look at what's going on. In particular she looks at where she was at a certain date and wonders if she was near Everest.*
S: Why don't you ask D [their editor]. He knows her personally.
*C carries on looking on Twitter.*
C: She's only done 127 tweets. So that's not very often, but she has tweeted in the last few days
*C sends D an email to see if he can find out anything more*

This example neatly emphasizes that whilst there may be a sense of dashboards servicing journalists independently, dashboards will typically be used in environments that are already rich in technological and human support. This has implications for how any kind of technological resource may get drawn upon, but also means that any UGC arriving via a dashboard is going to be just as subject to local discussion and assessment as any other potential lead. Thus dashboard use may be subject to both interactional and situational contingencies that may directly impact its value as a resource. If, as we have already noted, it cannot span device ecologies with those devices being potentially situated in different locations, if content cannot be easily shared in appropriate ways with colleagues, put aside, resurfaced, extracted and expanded, edited and annotated, and the myriad other things that already happen to possible story leads, then issues will surely arise.

*The audience:* Most journalists have a sense of their readership; a set of expectations regarding what their readers will want to read and how stories should be appropriately constructed and presented. The kinds of rigor devoted to fact checking can ride upon the extent to which readers are understood to care about this. The ways in which verification is made manifest within a story itself may also vary along these lines. UGC does not stand outside of these audience understandings and expectations. In the following example a sub-editor explains the implications of having an international readership that is largely made up of ex-pats:

Generally we assume we have an international audience. The Swiss abroad are a more natural target. So we are notionally delivering to an international audience but we are particularly aware of the Swiss audience.

So just what different journalists may wish to see in their own dashboard is going to vary at an organizational level.

*Story formats:* There are a variety of different news formats to which journalists may be writing and any one journalist may be working in several different formats during any single working day. These can span a trajectory from breaking news, to daily news, to feature articles. All of these may draw upon UGC in a variety of ways. Features may potentially examine the UGC surrounding a story in detail and present a broad analysis as a consequence. In breaking news for major events, individual tweets may be used verbatim and even embedded in a live stream. Different formats are also bound up with different kinds of time-scales. Breaking news requires rapid turnaround. Daily news is still very time sensitive but allows a little more time for reflection and analysis. Features may be developed over days or weeks. The nature of the format and associated time pressures also carry implications for the kinds of verification possible and this itself has import for the degree to which UGC may be considered viable.

In the case of breaking news there is an emphasis upon getting whatever information is available published and then updating stories as more information becomes available. In these kinds of cases journalists were even seen to enter stories directly into their online publishing tool rather than crafting them in Word first, which is their usual practice. In situations like this it is assumed that the story will be continually updated. Thus resources are revisited on a regular basis to examine whether anything has changed or there are new details. It is also the case here that *verification* of the details is itself treated as ongoing, with a tendency to avoid including details that have not been articulated elsewhere by 'trusted' organizations, and the use of safety-net phrases such as 'reports are coming in that …' rather than concrete assertions. In contrast, when journalists are working on features they may have several days to write their story and may schedule interviews with people ahead of time. In situations like this journalists are able to conduct more extensive research and assemble a wider array of materials before anything gets published than would be the case when working on the newsdesk.

Comparison of these two extremes makes it clear that just how UGC might be drawn upon in these different circumstances is divergent in a number of ways, both in terms of how exhaustively it might be explored but also in terms of how it might get incorporated into a story. To support such divergent uses of the same material is clearly a challenge for any dashboard. One way to tackle this is to focus instead upon some highly discrete aspect of the overall workflow (see, for instance, the focus on image forensics in [57]). However, an alternative approach, and the one we have attempted to follow, is to design something that is flexible enough to be able to encompass a variety of uses by building in an appreciation of the dimensions across which contingency may be seen to occur.

## CONCLUSIONS

One of the key problems that confront dashboard design as a consequence of the variability of both UGC use and the diverse ways in which journalists may need to verify the content it provides, is the capacity of any one dashboard to cover such a divergent set of requirements. Most dashboards assume that journalists will have the time to use them 'properly' and the capacity to drill into features that mark them out as unique in the services they aim to provide. They also presume that their *raison d'être* from the point of view of the designers is what journalists will actually seek to use them for in the context of live news production. However, the studies presented here have shown that for much of the time live news coverage is just too volatile for any resource to be drawn upon in anything more than a superficial fashion. This is the case until and when specific focus upon very specific features is occasioned. Indeed, our own evaluations showed that the initial dashboard design was better suited to the production of features because feature writing allows journalists time to *explore* data. This is a critical point. Most dashboards present some particular view upon some particular body of data in ways that will enable users to then explore that data and extract from it useful information. But, in the hothouse of competitive online news publication, journalists need their tools to surface the very thing they are looking for just now and to do it quickly. The trouble is knowing just what they might be looking for at any particular moment when the calculus is so complicated.

The work we have presented in this paper provides some scope for scaffolding this problem by enabling tools to be sensitive to the different ways in which variability in news production may be encountered. We are looking at how to design a dashboard that can surface immediately usable content quickly, but which, given time, can also provide a wider range of resources to service more detailed analysis. A number of features are also being built in that will enable journalists to structure the presentation of tweets and clusters of tweets in a variety of ways. We have particularly focused upon either enhancing or adding features that will provide newsdesk journalists with the capacity to identify pertinent information quickly in ways that are tailored to the specific current need. Key to this has been providing effective ways for journalists to search, sort and filter the information presented in step with unfolding content as it is posted on Twitter. Emphasis has also been placed upon providing coherent ways of grouping data that reflects unfolding stories and trends in a manner that is intuitive to the journalists themselves rather than locked in opaque machine logic. The latter concern in particular presents enormous challenges for dashboard design and there is an ongoing effort being devoted to overcoming it. Finally, for future work there is the equally difficult challenge of understanding how to make the decisions of complex machine learning tools accountable to their users [44, 56].

## ACKNOWLEDGMENTS
The research reported in this paper is supported by the EC FP7-ICT Collaborative Project PHEME (No. 611233).